\documentclass{skaox2006}
\usepackage{epsfig,graphics}
\def\grtsim{\mathrel{\hbox{\rlap{\hbox{\lower2pt\hbox{$\sim$}}}\raise2pt\hbox{$>$}}}}
\def\lesssim{\mathrel{\hbox{\rlap{\hbox{\lower2pt\hbox{$\sim$}}}\raise2pt\hbox{$<$}}}}
\begin{document}
   \title{A new method of detecting high-redshift clusters}

   \author{Caroline van Breukelen\inst{1}, Lee Clewley\inst{1}, David
          Bonfield\inst{1}}

   \institute{1. Astrophysics, Department of Physics, Keble Road, Oxford, OX1 3RH, UK}

   \abstract{We present a new cluster-finding algorithm based on a
     combination of the Voronoi Tessellation and Friends-Of-Friends
     methods. The algorithm utilises probability distribution
     functions derived from a photometric redshift analysis. We test
     our algorithm on a set of simulated cluster-catalogues and have
     published elsewhere its employment on UKIDSS Ultra Deep Survey
     infrared $J$ and $K$ data combined with 3.6 $\mu$m and 4.5 $\mu$m
     Spitzer bands and optical $BVRi'z'$ imaging from the Subaru
     Telescope. This pilot study has detected clusters over 0.5 square
     degrees in the Subaru XMM-Newton Deep Field. The resulting
     cluster catalogue contains 13 clusters at redshifts 0.61 $\leq z
     \leq$ 1.39 with luminosities $10 {\rm L^*} \lesssim L_{\rm tot}
     \lesssim 50 \rm L^*$.}

   \authorrunning{Caroline van Breukelen et al.}
   \maketitle
%
%
\section{Introduction}
Clusters of galaxies are the largest virialized structures in the
Universe and play an important role in our understanding of how
dark-matter haloes collapse and large-scale structure evolves. Their
number density can place constraints on the mass density of the
universe and the amplitude of the mass fluctuations (e.g. Eke et
al. 1998). Clusters also act as astrophysical laboratories for
understanding the formation and evolution of galaxies and their
environments. For instance the study of high-redshift clusters can
help us gain an understanding of the feedback processes caused by
star-formation and Active Galactic Nuclei (e.g. Silk \& Rees 1998). It
is therefore desirable to have a large, homogeneous catalogue of
clusters at a range of redshifts in the universe.

At present, however, there are only few clusters known at $z>1$ and
the vast majority of these are from X-ray surveys (e.g. using
XMM-Newton; Stanford et al. 2006). The paucity of known, distant
clusters stems from various selection effects or deficiencies. For
instance, optical searches for clusters, that work so efficiently at
$z<1$, are ineffective once the 4000 \AA\ break falls outside the
I-filter pass band ($z>1$) given the predominance of early-type, red
galaxies in clusters. One solution to this problem is to select
clusters in the near-infrared. Recent developments have seen the
advent of large near-infrared cameras like the Wide Field Infrared
Camera (WFCAM) on the United Kingdom Infrared Telescope (UKIRT). WFCAM
is now undertaking the UKIRT Infrared Deep Sky Survey (UKIDSS,
Lawrence et al. 2006), which is a suite of large area and deep
near-infrared sky surveys. UKIDSS provides the ideal opportunity to
search for high-redshift clusters in the infrared wavelength
regime. The survey has a high efficiency as it provides over an order
of magnitude increase in survey speed over existing near-infrared
imagers.

There are numerous methods for detecting clusters in optical/infrared
imaging surveys. The problem is easier with spectroscopic redshifts,
but currently spectroscopic information for infrared-selected
galaxies is in short supply, time consuming, and impractical over
large areas. However, approximate redshifts can be calculated via
photometric redshift estimation.  Despite being a popular method,
remarkably little attention is paid to using photometric redshifts to
isolate clusters. Further, with the exceptions of Kim et al. (2002),
Goto et al. (2002), Bahcall et al. (2003) and Lopes at al. (2004), very
little work has been done to compare the various cluster detection
method that exist (for a review see Gal 2005).


\section{The Algorithm}
We have developed a new cluster-detection algorithm [see van Breukelen
et al. (in preparation) for full details] to deal with two common
problems of photometric selection methods: (i) projection effects of
fore- and background galaxies and (ii) determining the reality of
detected clusters. The former issue arises because photometric - as
opposed to spectroscopic - redshifts typically have errors of the
order of $\sigma \sim 0.1$; furthermore the photometric redshift
probability functions (z-PDFs) are often significantly non-Gaussian
and can for instance show double peaks. To address this problem, our
cluster-detection algorithm utilizes the full z-PDF instead of a
single best redshift-estimate with an associated error. The second
issue - the occurrence of spurious cluster-detections - is due to
selection biases inherent in any detection algorithm. We take this
effect into account by cross-correlating the output of two
substantially different cluster-detection methods. Simulations reveal
that this reduces the contamination of the cluster sample to chance
galaxy groupings.

The algorithm is divided into six steps, described in
more detail in the following subsections.
\begin{enumerate}
\item Determining z-PDFs for all galaxies in the field.
\item Creating 500 Monte-Carlo (MC) realisations of the three-dimensional
  galaxy distribution, based on the galaxy z-PDFs.
\item Dividing each MC-realisation into redshift slices of $\Delta z =
  0.05$ over the range $0.1 \leq z \leq 2.0$.
\item Detecting cluster candidates in each slice of all
  MC-realisations using independent Voronoi Tessellation (VT) and
  Friends-Of-Friends (FOF) methods.
\item Mapping the probability of cluster candidates for both methods
  based on the number of MC-realisations in which they occur.
\item Cross-correlating the output of the VT and FOF methods to arrive
  at the final cluster-catalogue.
\end{enumerate}

\subsection{Photometric Redshifts}
We generate spectral energy distribution (SED) templates with the
stellar population synthesis code GALAXEV (Bruzual \& Charlot 2003),
which cover a range of different star formation rates with timescales
$\tau$ from 0.1 to 30 Gyr. Subsequently the redshift probability
functions are derived by fitting the SEDs to each galaxy's photometry
using the \emph{Hyperz} code (Bolzonella et al 2000). To create
marginalised posterior redshift probability functions we adopt a flat
prior for galaxy luminosity up to a maximum of $L = 10 \rm L^*$
(assuming a passively evolving elliptical galaxy) in the observed
$K$-band. 

\subsection{The Monte-Carlo realisations and redshifts slicing}
We create 500 Monte-Carlo realisations of the three-dimensional galaxy
distribution by randomly sampling each z-PDF. We then divide each
MC-realisation into redshift slices of $\Delta z = 0.05$. The width of
these slices is comparable to the error of the photometric redshift
error ($\sigma_z$); if it is chosen to be too small, clusters can be
undetected due to the distribution of their member galaxies over too
many redshift slices; if it is too large, many spurious sources will
be found owing to projection effects.

\subsection{Detection methods: VT and FOF}
Our algorithm applies the VT technique and FOF method independently to
each redshift slice of all the MC-realisations.

The VT technique divides a field of galaxies into Voronoi Cells, each
containing one object: the nucleus. All points that are closer to this
nucleus than any of the other nuclei are enclosed by the Voronoi
Cell. This technique was first applied to the modelling of large-scale
structure (e.g. Icke \& van de Weygaert 1987) but has more recently
been used in cluster detection (Ebeling \& Wiedenmann 1993; Kim et
al. 2002; Lopes et al. 2004). One of the principal advantages of the
VT method is that the technique is relatively unbiased as it does not
look for a particular source geometry (e.g Ramella 2001). The
parameter of interest is the area of the VT cells, the reciprocal of
which translates to a density. Overdense regions in the plane are
found by fitting a function to the density distribution of all VT
cells in the field; cluster candidates are the groups of cells of a
significantly higher density than the mean background density.

We follow the method of Ebeling \& Wiedenmann (1993); to determine the
mean background density of the VT cells we fit the following
cumulative function to the density distribution:

\begin{equation}
P(\tilde f) = e^{-4\tilde f} \Bigl (\frac{32}{3\tilde f^3} +
\frac{8}{\tilde f^2}+ \frac{4}{\tilde f} + 1 \Bigr ).
\end{equation}

Here $\tilde f$ is the cell density (the inverse of the cell area) in
units of the mean cell density, which is the parameter to be derived
from the fitting procedure. Once the background density is known, we
isolate all cells with $\tilde f > 1.75$. Adjoining high-density cells
are grouped; the groups that have a $> 90\%$ chance of not being a
background fluctuation (see Ebeling \& Wiedenmann, Section D) are
cluster candidates.

\begin{figure}
\begin{center}
\includegraphics[height=7cm]{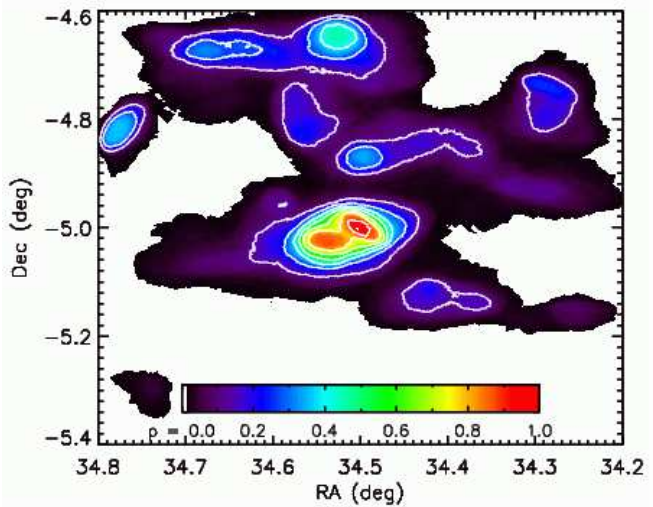}
\vspace{-3mm}
\caption{A probability map of clusters found by the Voronoi
  Tesselation method at redshift $z \sim 1.0$. Colours are normalised
  to the highest probability in the field.}
\label{contours}
\vspace{-5mm}
\end{center}
\end{figure}

\begin{figure*}
\begin{center}
\includegraphics[width=13cm]{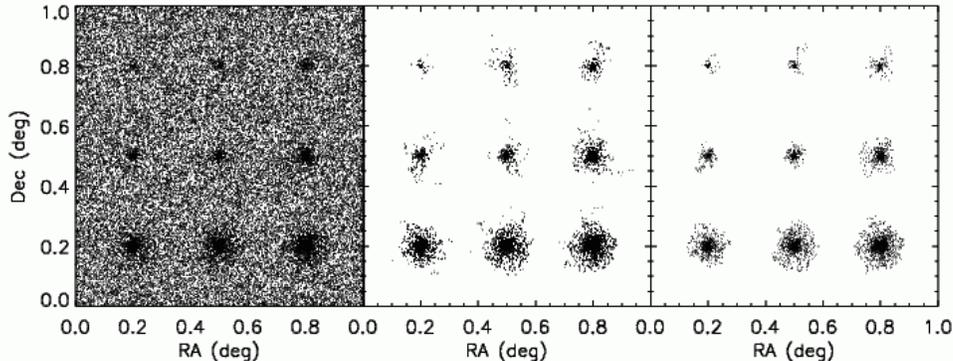}
\vspace{-3mm}
\caption{Example of simulated clusters superimposed on a galaxy
background (left) as recovered by the Voronoi tessellation technique
(middle) and the Friends-Of-Friends method (right). In this simulation
the clusters are spherical with a total luminosity of
10,20,30,40,50,100,150,200,300 $L^{*}$ (top left to bottom right
respectively) at $z=0.2$.}
\label{simulations}
\vspace{-5mm}
\end{center}
\end{figure*}

Friends-Of-Friends (FOF) algorithms are commonly used in spectroscopic
galaxy surveys (e.g. Tucker et al., 2002; Ramella et al., 2002). A
variant of this algorithm utilizing photometric redshifts was proposed
by Botzler et al. (2004). They create redshift slices for their data
cube and place the objects into the redshift slices according to their
photometric redshift and error. The algorithm then links galaxy pairs
within each slice that are closer to each other than some given
linking distance, $D_{\rm Link}$, which is the projected separation of
the galaxy pair. We use the empirically derived value of 0.175 Mpc for
$D_{\rm Link}$ (proper coordinates) and consider a minimum of five
galaxies in a group to be a cluster candidate.

An important difference between our algorithm and previous ones in the
literature is the way we place the galaxies in the redshift slices. As
we sample the full z-PDF to create MC-realisations of the
three-dimensional galaxy distribution, we do not need to assign errors
to individual galaxy redshifts. An object with a large redshift error
will be distributed throughout many different slices in the 500
MC-realisations, and therefore not yield a significant contribution to
the cluster candidates it is potentially found in. Thus there is no
need to remove objects with large errors from the catalogue and no
additional bias is introduced against faint objects with noisier
photometry. A second modification to existing algorithms is the way we
link up cluster candidates throughout the redshift slices. Instead of
comparing individual galaxies in the clusters and linking up the
clusters with corresponding members (see e.g. Botzler et al. 2004), we
use probability maps of all redshift slices to locate likely cluster
regions. This is discussed in the following section.

\subsection{Probability maps and cross-correlation}
Once the two cluster-detection methods have determined the cluster
candidates in the redshift slices for all MC-realisations, we combine
the MC-realisations to create a probability for both methods for each
redshift slice. Figure ~\ref{contours} shows an example of a
probability map: the VT cluster candidates in this slice at $z = 1.0$
are contoured and coloured, with black through to red indicating low
to high probability. This map is created by overplotting the extent of
all cluster detections; the regions of the field that are found to be
in a cluster in many MC-realisations are high-probability cluster
locations. Since the error on the photometric redshifts of the
galaxies is usually larger than the width of the redshift slices, each
cluster candidate is typically found in several adjoining slices. We
join the cluster candidates that occur in the same location in several
slices by locating the peaks in the probability maps and inspecting
the area within their contours in the adjoining redshift slices for
cluster candidates. All the cluster candidates found in the same
region in adjoining redshift slices are linked up into one final
cluster; the final cluster redshift is determined by taking the mean
of the redshift slices, weighted by the number of corresponding
MC-realisations. We assign a reliability factor $F$ to each cluster by
counting the total number of MC-realisations in which it occurs and
dividing it by the total of 500 realisations. We then cross-correlate
the cluster candidates output by the VT and FOF methods and take all
cluster candidates that are detected in both with $F > 0.2$. This is
the final cluster sample.

\section{Simulations}
\begin{figure*}
\includegraphics[height=52mm]{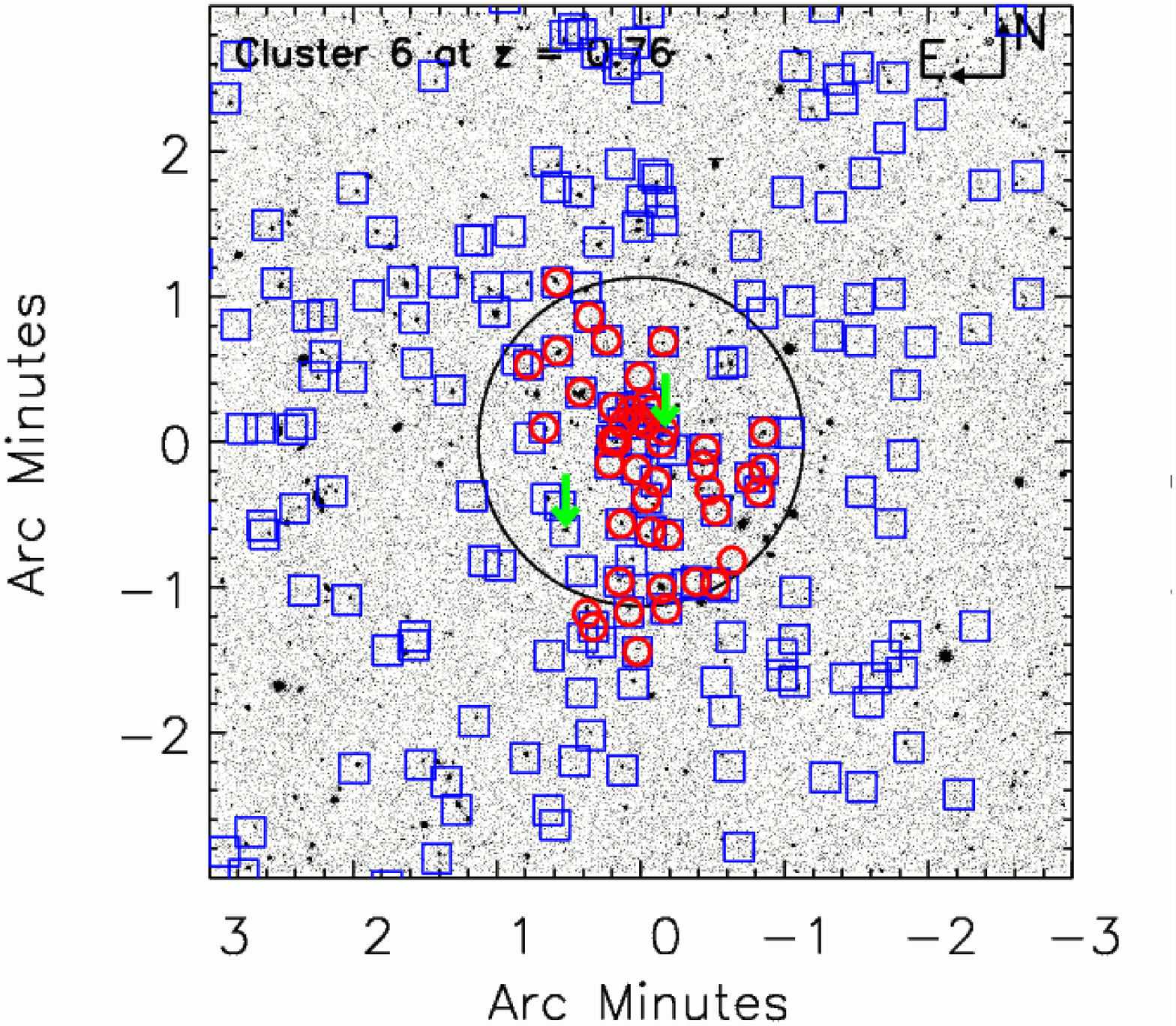}
\includegraphics[height=52mm]{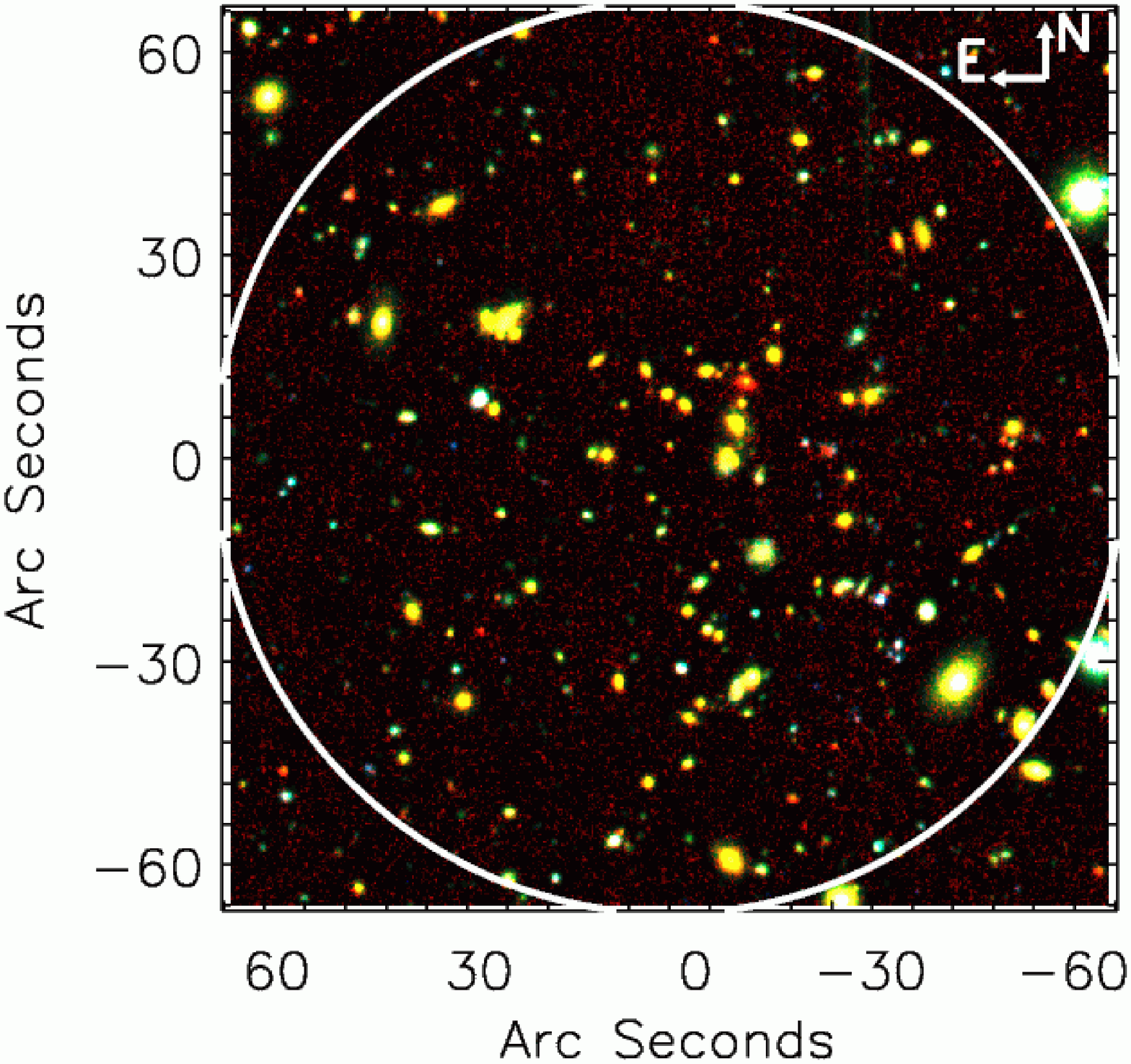}
\hspace{-6mm}
\includegraphics[height=56mm]{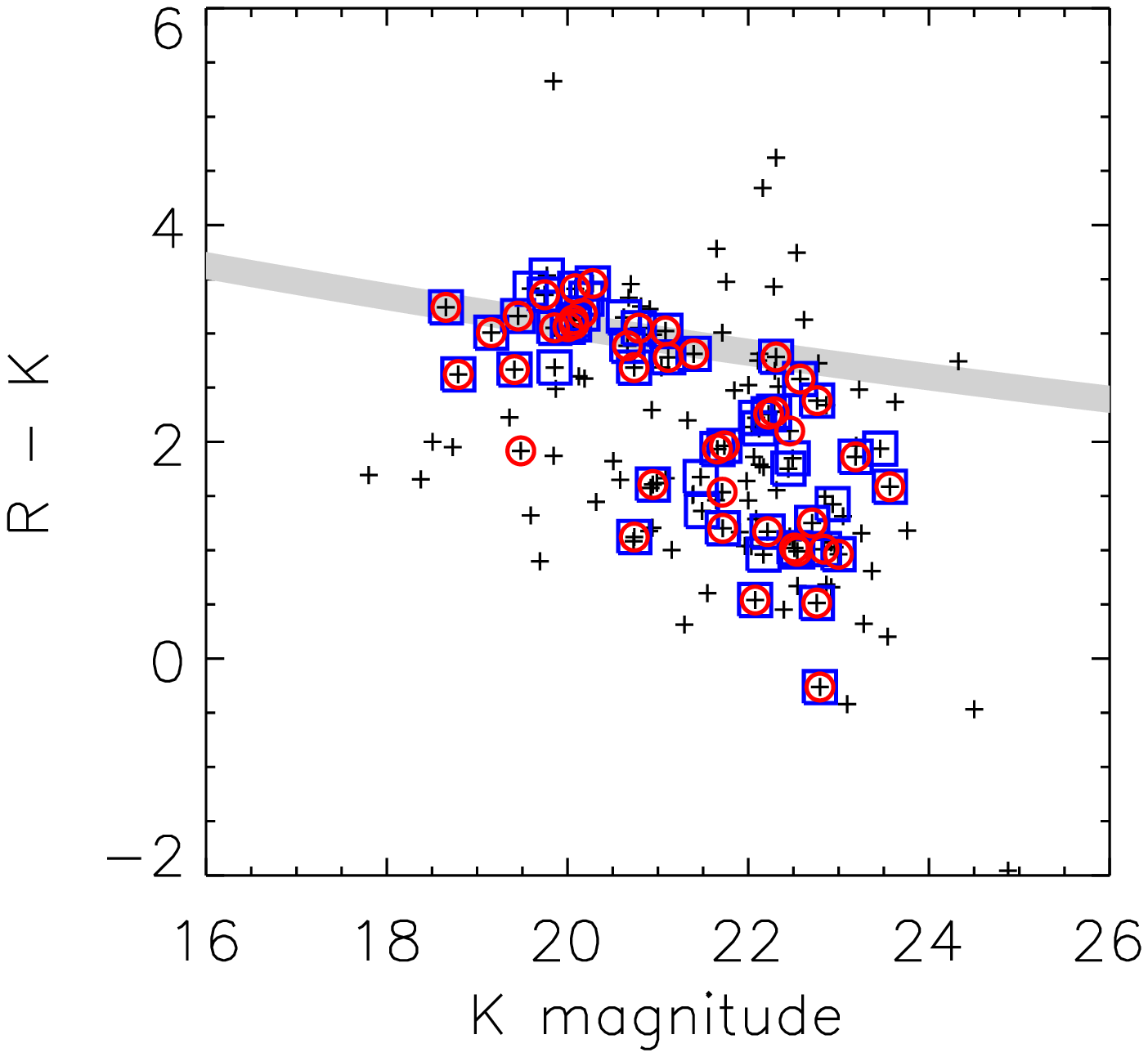}
\vspace{-3mm}
\caption{\small Cluster at $z=0.8$ detected with our cluster-finding
  algorithm from van Breukelen et al. (2006). {\em Left:} $K$-band image
  (UKIDSS-UDS); the large circle shows a 1 Mpc region around the
  cluster; the (blue) squares and (red) circles are cluster members as
  given by VT and FOF respectively. The (green) arrows point out
  galaxies with $z_{\rm spec} = 0.87$. {\em Middle:} $Bz'K$ image of
  the central 1 Mpc region. {\em Right:} colour-magnitude plot. The
  crosses are all galaxies within the central 1 Mpc region, otherwise
  the symbols are the same as in the left-hand panel. The grey band is
  the modelled red sequence.}
\label{colour_plots}
\vspace{-3mm}
\end{figure*}
To test the behaviour of the cluster-detection algorithm we run a set
of simulations on mock-catalogues. These comprise ten random versions
of clusters ranging in total luminosity from 10 $\rm L^*$ to 300 $\rm
L^*$ and redshift $0.1 < z < 2.0$, superimposed on a galaxy
distribution randomly placed in the field within the same redshift
range. Realistic galaxy luminosities and number densities are
determined by the $K$-band luminosity function of Cole et al. (2001)
for the field-distribution and Lin et al. (2004) for the clusters,
with the simplifying assumption of passive evolution with formation
redshift $z_{form}=10$. A detection limit of $K_{lim, AB} = 22.5$ is
imposed to match the 5-$\sigma$ limit of the UDS EDR data (see section
4). The galaxies are spatially distributed within a cluster according
to an NFW profile (Navarro, Frenk \& White, 1997) with a cut-off
radius of 1 Mpc. Figure~\ref{simulations} is an illustration of the
detection of simulated clusters by the VT and FOF methods. It shows an
example of a mock cluster catalogue with galaxy background (left) and
the clusters as recovered by VT (middle) and FOF (right). The smallest
cluster has a total luminosity of $L =10L^*$ and is detected by both
methods.

When comparing the recovered cluster galaxies to the input mock
cluster members, we see that VT tends to include all galaxies in a
large area around the cluster core and the number of recovered cluster
members, $N_{\rm gal, VT}$, is sensitive to the local field
density. By contrast, the galaxy members recovered by FOF are more
centrally concentrated and $N_{\rm gal, FOF}$ is consistent for mock
clusters of the same total luminosity throughout the random
realisations of the catalogues. Thus we can relate $N_{\rm gal, FOF}$
to the total luminosity of a detected cluster. We calculate $N_{\rm
gal, FOF}$ by taking all galaxies that occur in the cluster in $>15\%$
of the MC-realisations in which the cluster itself is detected. The
galaxies that appear in a smaller fraction of MC-realisations are very
likely to be interlopers from different redshifts. Once we know
$N_{\rm gal, FOF}$ for all simulated luminosities and redshifts, we
derive functions of $N_{\rm gal}$ vs. $z$ for constant total cluster
luminosity.

\section{Application to UKIDSS-UDS data}
To apply our cluster-detection algorithm to real observations, we used
three sources of data in our published paper (van Breukelen et
al. 2006): near-infrared $J$ and $K$ data from the UKIDSS Ultra Deep
Survey Early Data Release (UDS EDR, Foucaud et al. 2006); 3.6$\mu$m
and 4.5$\mu$m bands from the Spitzer Wide-area InfraRed Extragalactic
survey (SWIRE, Lonsdale et al. 2005); and optical $BVRi'z'$ Subaru
data over the Subaru XMM-Newton Deep Field (SXDF, Furusawa et al. in
prep.). In this pilot study we restricted ourselves to a rectangular
area of 0.5 square degrees, exhibiting a survey-depth of $K_{\rm
AB,lim} = J_{\rm AB,lim} = 22.5$ (UDS EDR 5$\sigma$ magnitude
limits). We included objects with a detection in $i'$, $J$ and $K$ in
the galaxy catalogue and to exclude stars we impose a criterion of
SExtractor stellarity index $<$ 0.8 in $i'$ and $K$ (e.g. Bertin \&
Arnouts 1996). Subsequently we ran our photometric redshift code on
this sample, resulting in a redshift catalogue of 19300 objects in the
range $0.1 \leq z \leq 2.0$.

\subsection{Results: the UKIDSS-UDS cluster catalogue}
Application of our cluster-detection algorithm to the redshift
catalogue yielded 14 clusters at $0.61 \leq z \leq 1.39$ (van Breukelen
et al. 2006). Figure ~\ref{colour_plots} shows one of the detected
clusters: on the left a K-band image with the cluster members marked
(FOF: circles, VT: squares); in the middle a $Bz'K$ image of the
central 1 Mpc region of the cluster; on the right a colour-magnitude
diagram with the modelled red sequence overplotted.

To derive the clusters' luminosity, we compared the results to the
output of the simulations. We determined $N_{\rm gal, FOF}$ with $K <
22.5$ (corresponding to the completeness limit) in the same way as for
our simulated clusters; this allowed us to derive an approximate total
luminosity to the cluster by interpolating between the lines of
constant total luminosity in the $N_{\rm gal}-z$ plane found in our
simulations. We found our clusters span the range of $10 {\rm L^*}
\lesssim L_{\rm tot} \lesssim 50 \rm L^*$; assuming $\frac{M/\rm
M_{\odot}}{L/\rm L_{\odot}} = 75h$ (Rines et al. 2001) this yields
$0.5 \times 10^{14}~{\rm M_{\odot}} \lesssim M_{\rm cluster} \lesssim
3 \times 10^{14}~\rm M_{\odot}$. 

Clearly spectroscopic observations of these clusters are essential to
confirm their reality, particularly for the high-redshift clusters
which are cosmologically more valuable. In the near future highly
multiplexed multi-object spectrometers on 8-metre class telescopes
will provide the ideal opportunity for spectroscopical follow-up of
high-redshift clusters. X-ray data and radio observations of the
Sunyaev-Zel'dovich effect will also be able to convincingly confirm
the reality of the clusters.
\begin{acknowledgements}
We are grateful to our other collaborators on this project and
acknowledge funding from PPARC.
\end{acknowledgements}

\end{document}